\documentclass[a4paper]{jpconf}

\usepackage{amssymb,amstext}

\newcommand{\vn}{{\vec{n}}{}}
\newcommand{\itGamma}{\Gamma}
\newcommand{\G}{\itGamma}

\newcommand{\prirodni}{\ensuremath{\mathbb{N}}}
\newcommand{\ds}{\displaystyle}

\begin{document}

\title{Effective action for EPRL/FK spin foam models}

\author{Aleksandar Mikovi\'c$^{1,2}$ and Marko Vojinovi\'c$^{2}$}

\address{$^1$Departamento de Matem\'atica, Universidade Lus\'ofona de Humanidades e Tecnologias \\ $\,\,$ Av. do Campo Grande, 376, 1749-024 Lisboa, Portugal}
\address{$^2$Grupo de Fisica Matem\'atica da Universidade de Lisboa \\ $\,\,$ Av. Prof. Gama Pinto, 2, 1649-003 Lisboa, Portugal}

\ead{amikovic@ulusofona.pt, vmarko@cii.fc.ul.pt}

\begin{abstract}
We show that a natural modification of the EPRL/FK vertex amplitude gives a finite spin foam model whose effective action gives the Einstein-Hilbert action in the limit of large spins and arbitrarily fine spacetime triangulations. The first-order quantum corrections can be easily computed and we show how to calculate the higher-order corrections.
\end{abstract}

\section{Introduction}
The EPRL/FK spin foam models \cite{EPRL,FK} represent a class of spin foam models whose boundary states are the Loop Quantum Gravity spin networks, while the vertex amplitudes are constructed such that all the discretized constraints of the Plebanski action are imposed. The partition function for an EPRL/FK spin foam model can be written as
\begin{equation}
Z = \sum_j \int d\vn \prod_f W_f(j_f) \prod_v W_v(j_f,{\vn}_{lf} ) \,,\label{scos}
\end{equation}
where $j_f$ are $SU(2)$ spins assigned to the faces $f$ of the dual $2$-complex $K$ for a spacetime triangulation, $\vn_{lf}$ are the Livine-Speziale normal vectors, one for each edge $l$ and the adjacent face $f$ of $K$, see \cite{LivineSpeziale}. The face amplitude $W_f(j)$ can be chosen to be $2j_f+1$, see \cite{Bianchi}, while $W_v$ is the vertex amplitude, which can be represented as a certain linear combination of the Lorentz group $15j$-symbols, see \cite{EPRL,FK} for details.

The classical limit of the quantum gravity theory defined by the path integral (\ref{scos}) can be determined by computing the corresponding effective action. The effective action can be expanded as the classical action plus the quantum corrections, which are proportional to the powers of $\hbar$, so that in the limit $\hbar\to 0$ one will obtain the classical action. One way to do this is to calculate the graviton correlation functions in the large-spin limit, see \cite{grp,trpc}. However, this approach has a drawback, since it requires the knowledge of infinitely many correlation functions in order to construct the corresponding effective action.  On the other hand, the effective action can be computed directly \cite{MVeffaction}, which is much easier to do. We will sketch this computation here. 

In order to calculate the effective action, the partition function has to be finite. The convergence of (\ref{scos}) can be studied by introducing a real deformation parameter $p$ such that $Z \to Z_p$, where 
$$
W_v \to \frac{W_v}{\prod_{f\in v} \left( 2j_f + 1 \right)^p}\,,
$$
and $W_f$ stays the same, see \cite{MVfiniteness}. One also needs to know the large-spin asymptotics of $W_v$, which is given by
\begin{equation} \label{verta}
W_v (j,\vn) \approx  \frac{1}{V(j)}\left[ N_+(j) e^{i\alpha S_{vR}(j)} + N_-(j) e^{-i\alpha S_{vR}(j)}\right] \, ,
\end{equation}
where $V(j)$ is a homogeneous function of order 12, $N_\pm (j)$ are homogeneous functions of order zero, $\alpha=\gamma$ or $\alpha=1$, depending on whether the vertex boundary constitutes a Lorentzian or Euclidean Regge geometry, respectively. $\gamma$ is the Barbero-Immirzi parameter and $S_{vR}$ is the Regge action for the vertex $v$. If $(j,\vn)$ at a vertex do not form a Regge geometry, then $W_v (\lambda j,\vn)$ falls off faster that any power of $1/\lambda$, see \cite{Barrett,Fairbain}.

One also needs the asymptotics for the case when some of the $j$ are big and other are small, but it is not difficult to see that in any case $|W_v(j, \vn)|$ is a bounded function. Then it can be shown that $Z_p$ is absolutely convergent for $p>1$, see \cite{MVfiniteness}. Another complication which has to be resolved before calculating the effective action is that the asymptotics (\ref{verta}) is of the $\cos(S_R)$ type, instead of the $e^{iS_R}$ type. This can be resolved by redefining the vertex in the following way
\begin{equation} \label{redv}
W_v  \to  A_v = \frac{1}{\ds \prod_{f\in v} \left( 2j_f + 1 \right)^p} \frac{W_v + \sqrt{W_v^2 - 4N_+N_-/V^2}}{2N_+} \, .
\end{equation}
Then 
\begin{equation} \label{asredv}
A_v(j,\vn) \approx  \frac{1}{\ds \prod_{f\in v} \left( 2j_f + 1 \right)^p}\, \frac{e^{i\alpha S_{vR}(j)}}{V(j)} \, ,
\end{equation}
when all $j$ are large. 

If $N_+ \ne N_-$ one can obtain the exponential asymptotics of the vertex amplitude by making a linear combination of $W$ and its complex conjugate \cite{MVgpa}. Another way is to use the fact that $W$ is a linear combination of  $15j$-symbols, and to change the coefficients such that the exponential asymptotics is achieved, see \cite{engle}.

\section{Effective action}

An effective action $\G$ can be defined for a spin foam model by adapting the the background field method from quantum field theory (QFT), see \cite{MVeffaction}. In the case of the EPRL/FK models one obtains
\begin{equation} \label{bfmea}
e^{i\itGamma(j,\vn)} = \sum_{j'} \int d\vn' \prod_f W_f(j+j') \prod_v A_v(j+j',\vn + \vn') \, ,
\end{equation}
where $(j,\vn)$ are the background variables and $j' \in \prirodni/2$. This formula is valid in the semiclassical limit, i.e. when $j\to\infty$ and the Barbero-Immirzi parameter $\gamma$ is fixed. One can then employ the asymptotic formula (\ref{asredv}) for the vertex amplitude in the definition of the effective action (\ref{bfmea}), since the background spins $j$ are all large. It is known that the background configurations $(j,\vn)$ which form a Regge geometry have a dominant contribution to the path integral (\ref{scos}) for large spins, see \cite{FreidelConrady,MagliaroPerini}. In that case $(j,\vn)=(j(L), \vn(L))$, where $L$ is a set of lengths for the edges of the spacetime triangulation, and one can show by using (\ref{asredv}) and the standard Gaussian integral formulas, see \cite{MVeffaction}, that
\begin{equation} 
\itGamma(j,\vn) = S_R(L) +\sum_f c_f(p) \log j_f + \frac{1}{2} \Tr \log {\tilde S}_R''  + O(1) \, .\label{foea}
\end{equation}
$S_R(L)$ denotes the Regge action for the spacetime triangulation and $c_f (p) = 1 - (p+12) n_f$, where $n_f$ is the number of vertices of a face $f$. The matrix ${\tilde S}''_R$ is given by
$$
\tilde{S}''_R = \left(\matrix{\tilde{S}''_{R\,jj} \,\, S''_{R\,jn} \cr S''_{R\,jn} \,\, S''_{R\,nn}}\right)\,, 
$$
where $S''_{jj}$, $S''_{jn}$ and $S''_{nn}$ denote the matrices formed from the corresponding second-order partial derivatives of $S_R = \gamma \sum_f j_f \Theta_f (j , \vn)$ evaluated at $j=j(L)$ and $\vn=\vn(L)$. The elements of the matrix $\tilde{S}''_{jj}$ are given by
$$
\tilde S''_{R\,ff'} = S''_{R\,ff'} -i \frac{c_f}{j_f^2}\delta_{ff'} \,.
$$

The notation $O(1)$ in (\ref{foea}) means that
\begin{equation} \label{eaexp}
\G(\lambda j,\vn) = \lambda \tilde{\G}_0(j,\vn) + (\log\lambda) \tilde{\G}_1(j,\vn)  + \tilde{\G}_2 (j,\vn)+ \lambda^{-1} \tilde{\G}_3 (j,\vn) + \cdots\,,
\end{equation}
so that the first term in (\ref{foea}) is of $O(j)$, the second and the third term in (\ref{foea}) are of $O(\log j)$, the next term is of $O(1)$ and so on. Hence the leading term for $j\to\infty$ is the Regge action, which means that it is the classical limit of the theory. If the triangulation of the spacetime is fine enough, the Regge action can be approximated by the Einstein-Hilbert action as $S_R \approx S_{EH}/2$, and one recovers general relativity in the limit of smooth geometry \cite{MVeffaction}.

The second term in (\ref{foea}) comes from the face amplitude and the regularization factor in the vertex amplitude, while the third term is a discretization of the trace-log term from QFT.

\section{Higher-order corrections}

The method used in \cite{MVeffaction} to calculate the effective action was based on the ``on-shell'' background, i.e. the background $(j,\vn)$ was chosen such that $(j,\vn)$ formed an ``on-shell'' Regge geometry. In this case $\vn=\vn(L_0)$ and $j=j(L_0)$ where $L_0$ is a set of edge lengths which satisfy the Regge action equations of motion. After the calculation was done, the $L_0$ background was replaced with a background $L$, where the lengths in the set $L$ do not have to satisfy the Regge equations. 

In QFT it is known that such an approach is accurate only up to the first-order in $\hbar$, since the exact formula for the effective action is given by
\begin{equation}
e^{\frac{i}{\hbar}\G(\phi)} = \int {\cal D}h \exp\left(\frac{i}{\hbar} S(\phi + h) - \frac{i}{\hbar}\int dx\, \frac{\delta\G}{\delta\phi(x)}\,h(x)\right)\,,\label{qfteea}
\end{equation}
see \cite{nair}. The formula (\ref{qfteea}) is an integro-differential equation, which can be solved perturbatively as
\begin{equation}
\G = S + \hbar \G_1 + \hbar^2 \G_2 + \cdots \,,\label{hbare}
\end{equation}
and one does not need to use an on-shell background $\phi$ in the calculation.

The exact QFT formula (\ref{qfteea}) can be applied to an EPRL/FK spin foam model as
\begin{equation} \label{esfea}
e^{i\G(j,\vn)} = \sum_{j'} \int d\vn' \prod_f W_f(j+j')\,\prod_v {W_v}(j+j',\vn+\vn')\,e^{ -i \left\langle\frac{\partial \G}{\partial j}  j' + \frac{\partial \G}{\partial \vn}  \vn' \right\rangle}\, ,
\end{equation}
where
$$
\left\langle\frac{\partial \G}{\partial j}  j' + \frac{\partial \G}{\partial \vn} \vn' \right\rangle =\sum_f\frac{\partial \G}{\partial j_f}  j'_f + \sum_{fl}\frac{\partial \G}{\partial \vn_{fl}}  \vn'_{fl}\,,
$$
is the spin-foam analog of the higher-loop correction term from (\ref{qfteea}).

The background $(j,\vn)$ will be chosen to be a Regge background $(j(L),\vn(L))$, where the lengths $L$ are off-shell, i.e. do not have to satisfy the Regge equations of motion. If the background is chosen such that it does not form a Regge geometry, then we expect that $e^{i\G(j,\vn)}$ will be exponentially suppressed, i.e.
$$
\left| e^{i\G(j,\vn)}\right|_{j\ne j(L),\vn\ne \vn(L)} \ll \left| e^{i\G(j,\vn)}\right|_{j= j(L),\vn= \vn(L)} \,.
$$

The analog of the $\hbar$-expansion (\ref{hbare}) is the formula (\ref{eaexp}), which we will write as
$$
\G = S_R + \G_1 + \G_2 + \cdots + \G_n +\cdots ,
$$ 
where $S_R = O(j)$, $\G_1 = O(\log j)$ and $\G_n = O(j^{-n+2})$ for $n\ge 2$. By using the techniques from \cite{MVeffaction} it is not difficult to show that $\G_1$ is the same as the $O(\log j)$ term from (\ref{foea}), while the differences between the formula (\ref{bfmea}) and the formula (\ref{esfea}) will appear for $n\ge 2$ terms.

\section{Conclusions}

The effective action approach from QFT is quite useful for exploring the semiclassical limit of spin foam models of quantum gravity since the calculations in this approach are much easier than the calculations in the other approaches. The effective action approach can also tell us what kind of vertex amplitude we should choose in order to obtain a spin foam model whose classical limit is GR. 

The first-order quantum corrections have two terms, see (\ref{foea}). The third term in (\ref{foea}) is a discrete analog of the usual trace-log term from QFT, while the second term is specific for spin foam models. It will be important to find out what is the form of these quantum corrections in the smooth geometry limit, since these corrections will determine the first-order back-reaction effects for black holes and the cosmological evolution. The higher-order corrections can be calculated from the equation (\ref{esfea}). 

It would be interesting to apply the effective action approach to the Regge model, and compare the results to the results obtained in the spin foam case.

\ack

This work has been partially supported by FCT project PTDC/MAT/099880/2008. MV was also supported by the FCT grant SFRH/BPD/46376/2008.



\end{document}